\documentstyle[aps]{revtex}

\begin{document}
\author{Mario Castagnino}
\title{The equilibrium limit of the Casati-Prosen model}
\address{Institutos de Astronom\'{i}a y F\'{i}sica del Espacio y de F\'{i}sica\\
Rosario.\\
Casilla de Correos 67 Sucursal 28\\
1428 Buenos Aires, Argentina.\\
e-mail: mariocastagnino@citynet.net.ar}
\maketitle

\begin{abstract}
An alternative explanation of the decoherence in the Casati-Prosen model is
presented. It is based on the Self Induced Decoherence formalism extended to
non-integrable systems.

PACS number 0365 Yz

Key works: decoherence, interferences, billiards, slits, quantum chaos.
\end{abstract}

\section{Introduction.}

The Casati-Prosen model \cite{CP} combines two paradigmatic models of
classical and quantum mechanics: a Sinai billiard, where the simplest
examples of chaotic motion take place and a Young, two slits, experiment,
the main example of quantum behavior that it ''...is impossible absolutely
impossible to explain in any classical way''\cite{Feynman}. So we really
could call this model the ''Sinai-Young'' experiment. We consider that the
complete understanding of this model is essential to solve problems like
quantum irreversibility, decoherence, and chaos. The model is shown in
figure 1 (of paper \cite{CP}), namely a triangular {\it upper billiard} with
perfectly reflecting layers, with two slices in its base, on the top of a
box, the {\it radiating region}, with a photographic film in its base and
absorbent walls. A quantum state with a gaussian packet initial condition
bounces in the triangle, and produces two centers of radiation in the two
slices from which a small amount of probability current leaks from the
billiard to the radiating zone. Then when the billiard is perfectly
triangular and therefore integrable (full-lines in figure 1 of \cite{CP})
the interference fringes (full-lines in figure 2 ) appear in the film and
when it is a Sinai billiard and therefore non-integrable (dotted line of fig
1) the first pattern decoheres to the (dotted) curve of figure 2. This
computer experiment shows how complexity can produce decoherence (without an
environment or an external noise) and it is explained in paper \cite{CP}
using a kinematical average. As the subject is so important we would like to
add another feature to the Casati and Prosen explanation of the phenomenon
showing that the model reaches an equilibrium state where decoherence
appears. In doing so we will use our previous results on decoherence \cite
{SID}, mainly paper \cite{ChSF}, where local constants of the motion are
introduced both at the classical and quantum level allowing to define
non-integrable quantum systems and to give a minimal definition of quantum
chaos, and paper \cite{DT}, where decoherence times are found. Recently we
have shown \cite{SIDEID} that our formalism, ''Self Induced Decoherence''
(SID) can be encompassed with the traditional one ''Environment Induced
Decoherence'' (EID) \cite{EID}, combining the advantages of both formalisms.

\section{The problems of paper \protect\cite{CP}.}

To make clear our physical point of view let us consider the two main
problems to understand chaotic motion in terms of quantum mechanics listed
in the introduction of paper \cite{CP} (see also \cite{5}):

1.- How is it possible to find chaos in bound systems, with finite number of
particles, which have a quasi-periodic behavior and therefore a discrete
evolution spectrum, if chaotic (e. g. mixing) motion requires a continuous
one?

We consider that the solution can be found in paper \cite{Alimanias} where
it is shown that, even if a quantum system has a discrete evolution
spectrum, the motion can be modeled with a continuous spectrum for times
much smaller than recurrence or Poincar\'{e} time. For a discrete energy
spectrum $\{\alpha _{\nu }\}$ this time is 
\begin{equation}
t_{P}\approx \frac{2\pi \hbar }{\min (\alpha _{\nu +1}-\alpha _{\nu })}
\label{2-1}
\end{equation}
so if the distances among the eigenvalues are very small $t_{P}$ is
extremely large. Then for $t\ll t_{P}$ the typical theorems, e.g. the
Riemann-Lebesgue theorem, can be used.

2.- In quantum motions initial errors propagate linearly while in chaotic
system this propagation is exponential. This contradiction makes quantum
chaos impossible.

We consider that, most likely, this kind of reasonings is done in quantum
systems with an integrable classical system as classical limit. If this is
not the case (as in the systems studied in \cite{ChSF}) it can be
demonstrated that the trajectories in the classical limit are chaotic and
may have positive Lyapunov exponents. So the contradiction is solved.

\section{The bases of our alternative explanation.}

We will give our alternative explanation based in three results. In this
section we only give a sketch of the main ideas on these subjects, the
complete treatment and figures can be found in the references:

a.- In paper \cite{ChSF}, using the Weyl-Wigner-Moyal isomorphs the
definition of classical integrable and non-integrable system is extended to
the quantum case. Then the SID formalism is extended to the non-integrable
system. For $N$ configuration variables these systems, in the classical case
have less than $N$ global constants of the motion. But according to the
Carath\'{e}orory-Jacobi lemma \cite{CJ} they have $N$ constants of the
motion locally defined which, via a Weyl-Wigner-Moyal isomorphism, allow to
define $N$ local Complete System of Commuting Observables that are used in
the extension of SID. The resulting theory is very similar to the original
one. Only an extra index $i$ corresponding to the domain $D_{\phi _{i}}$
(that contains the point of the phase space $\phi _{i}$ and where local
constant of motion are defined) must be added in all summations. Then the
state of the system $\rho (t)$ reaches an equilibrium state $\rho _{*},$
given by eq. (3.23) of \cite{ChSF}, defined as a weak limit 
\begin{equation}
W\lim_{t\rightarrow \infty }\rho (t)=\rho _{*}=\sum_{imm^{\prime
}}\int_{0}^{\infty }d\omega \rho (\omega )_{\phi _{i}p}(\omega ,m,m^{\prime
}|_{\phi _{i}}  \label{4-1}
\end{equation}
where $\omega $ is the eigenvalue of the hamiltonian $H$ (which is
considered to be globally defined), $m_{\phi _{i}}{\bf =(}m_{x},m_{y})_{\phi
_{i}}$ in our case will be the eigenvalue of the local momentum ${\bf %
P_{\phi _{i}}=(}P_{x},P_{y})_{\phi _{i}},$ and $(\omega ,m,m^{\prime
}|_{\phi _{i}}$ the cobasis of the eigen basis of the CSCO \{$H,{\bf P_{\phi
_{i}}\},}$ namely \{$|\omega ,m\rangle _{\phi _{i}}\langle \omega ,m^{\prime
}|_{\phi _{i}}\}$. Then the equilibrium final state $\rho _{*}$ has
decohered in the energy since only the diagonal terms in $\omega $ appear
(if not the basis would be $(\omega ,\omega ^{\prime },m,m^{\prime }|_{\phi
_{i}})$ but not in the remaining observables ${\bf (}P_{x},P_{y})_{\phi
_{i}} $, since non-diagonal terms $m,$ $m^{\prime }$ do appear. Then via a
simple diagonalization in the indices $m,m^{\prime }$ we reach to eq. (3.33) 
\begin{equation}
W\lim_{t\rightarrow \infty }\rho (t)=\rho _{*}=\sum_{ip}\int_{0}^{\infty
}d\omega \rho (\omega )_{\phi _{i}p}(\omega ,p,p|_{\phi _{i}}  \label{4-2}
\end{equation}
where $p_{\phi _{i}}{\bf =(}p_{x},p_{y})_{\phi _{i}}$ are the eigenvalues of
an adequate CSCO \{$H,{\bf O_{\phi _{i}}\}.}$ In the correspondent
eigenbasis $\rho _{*}$ is fully decohered, since now only diagonal terms (in 
$\omega $ and $p)$ appear.

b.- The upper triangle will be considered as the Sinai billiard of appendix
A of paper \cite{ChSF}. Namely the triangle will be complemented by three
potential walls in such a way that these potentials $U(x,y)$ (similarly to
those of the Sinai billiard of appendix A of paper \cite{ChSF}) produce the
bounces against the sides of the triangle. We will call $D_{0}$ the interior
of the triangle (as the $D_{0}$ of figure 2 of \cite{ChSF}). It has two
independent local constants of the motion: $H$ and $P_{x}$ (or $H$ and $%
P_{y} $ or $P_{x}$and $P_{y}$ since $H=\frac{1}{2M}($ $P_{x}^{2}+$ $%
P_{y}^{2})).$ Then we will add three extra domains, each one for each
potential wall, $D_{1},D_{2},D_{4}.$ In the case of the triangle with
straight sides the local constants of the motion in the boundaries can be
deduced by their symmetries. They are:

\begin{itemize}
\item  In the horizontal boundary ($U(x,y)=U_{1}(y),$ domain $D_{1})$ $H$
and $P_{x}.$

\item  In the vertical boundary ($U(x,y)=U_{2}(x),$ domain $D_{2})$ $H$ and $%
P_{y}$

\item  In the third boundary ($U(x,y)=U_{4}(ax+yb)$ domain $D_{4})$ $H$ and
a linear combination of $P_{x}$ and $P_{y}.$
\end{itemize}

This is not the case if the triangle has a circular boundary (with radius $%
r=a$ and angular coordinate $\theta ,$where the constants of the motion in
the third boundary ($U_{4}(x,y)=U(r),$ domain $D_{4})$ are $H$ and $%
P_{\theta }.$

c.- Decoherence times $t_{D}$ will be calculated using references \cite{DT}
and \cite{N}. From reference \cite{DT} we know that 
\[
t_{D}=\frac{\hbar }{\gamma } 
\]
where $\gamma $ is the distance to the real axis of the pole of the
resolvent closer to this axis. These poles for a circular symmetric
potential can be computed from reference \cite{N} From eqs. (5.1.4) and
(5.5.24) of this reference we know that the energy is 
\[
E=\frac{\hbar ^{2}k^{2}}{2M},\text{ and }k=\frac{\beta }{a} 
\]
where $M$ is the mass, being, from eq. (5.5.29), the $\beta $ for the pole
closer to the real axis 
\[
\beta _{0}=R_{0}-iI_{0}=U_{0}-\left( \frac{m+2}{4U_{0}}\right) \ln \left( 
\frac{2U_{0}^{m+2}}{A^{2}}\right) -\frac{i}{2}\ln \left( \frac{2U_{0}^{m+2}}{%
A^{2}}\right) 
\]
where $U^{(m)}(a-)$ is the first non vanishing derivative of the potential
at the boundary (corresponding to the side of the potential) and
coefficients $U_{0}$ and $A$ are given by eqs. (5.5.24) and (5.5.26) of \cite
{N}.

Then 
\begin{equation}
\gamma =\frac{\hbar ^{2}R_{0}I_{0}}{2Ma^{2}},\text{ and }t_{D}=\frac{2Ma^{2}%
}{\hbar R_{0}I_{0}}  \label{4-3}
\end{equation}
Below we will use this equation.

\section{The triangle with straight sides.}

Let us first consider the case of the straight triangle and let us take as
initial condition in the triangle a pure state wave packet $|\varphi \rangle
\sim |\varphi ({\bf x,}0)\rangle $ (which of course it is not an eigenstate
of the momentum operator ${\bf P}$). With this initial condition we obtain
the solution $|\varphi \rangle \sim |\varphi ({\bf x,}t)\rangle $ in the
triangle$,$ that can be written as a matrix 
\begin{equation}
\rho (t)=|\varphi ({\bf x,}t)\rangle \langle \varphi ({\bf x,}t)|
\label{CP.1}
\end{equation}
We can make some remarks:

i.- If the billiard is considered classical, two initial parallel
trajectories remain parallel while they bounce in the triangle. Therefore
there are neither positive Lyapunov exponent nor chaos.

ii.- Even if according to eqs. (\ref{4-1}) or (\ref{4-2}) there will be
decoherence in an infinite time, in this case the characteristic decoherence
time is, in fact, infinite since all the potential walls in this case are
straight lines and therefore the radius $a\rightarrow \infty $, then from
eq. (\ref{4-3}) $t_{D}\rightarrow \infty $ . Therefore $\rho (t)$ remains
bouncing forever  in the triangle and does not decohere.

Let us now consider the lower part under the slit screen.. The direct impact
of the packet (\ref{CP.1}) produces two boundary conditions in the two slits:%
$.$ These two boundary conditions produce two circular-symmetric solutions, $%
|\varphi _{1}({\bf x,}t)\rangle $ and $|\varphi _{2}({\bf x,}t)\rangle $,
with centers of symmetry in the two slits$.$ Therefore the state in the
lower part is $|\varphi ({\bf x,}t)\rangle $ $=|\varphi _{1}({\bf x,}%
t)\rangle $ $+$ $|\varphi _{2}({\bf x,}t)\rangle $ and the probability at $%
{\bf x}$ is${\bf :}$%
\[
p=\langle |{\bf x\rangle \langle x|\rangle }_{|\varphi ({\bf x,}t)\rangle
\langle \varphi ({\bf x,}t)|}=p_{1}+p_{2}+p_{int}
\]
where 
\begin{equation}
p_{1}=|\varphi _{1}({\bf x,}t)|^{2}\geq 0,\text{ }p_{2}=|\varphi _{2}({\bf x,%
}t)|^{2}\geq 0,\quad p_{int}=2%
\mathop{\rm Re}%
(\varphi _{1}({\bf x,}t)\varphi _{2}^{*}({\bf x,}t))  \label{CP.2}
\end{equation}
Of course $p_{int}\neq 0$ and it is the interference term. Let us observe
that as $|\varphi ({\bf x,}t)|^{2}$ is time invariant $p$ is also time
invariant as it is verified in \cite{CP} figure 4a.

Moreover if we consider many bounces of the packet instead of just the
direct impact, instead of (\ref{CP.1}) we will have a sum with different
momenta ${\bf P}$. But if the system is integrable this sum will have a
finite number of terms (see \cite{CP} and \cite{8}) and the interference
fringes will remain.

\section{The triangle with a curved side.}

Let us now consider the case of the curved triangle. Now

i.- Initial parallel trajectories will lose their parallelism when they
collide with the curved side and there will be positive Lyapunov exponent
and chaos. In fact, Sinai billiards are K-systems.

ii.- The potential walls are not trivial (i.e. $a\neq \infty $) and
therefore the analytic continuation of the resolvent has complex poles and
the finite decoherence time is given by eq. (\ref{4-3})\footnote{%
Essentially, as explained, the system has a central domain $D_{0}$ and three
potential boundaries domains $D_{1},$ $D_{2},$ and $D_{4}$. But each
scattering in the potential of $D_{4}$ can be considered as beginning in the
domain $D_{0}^{-}=D_{\phi _{1}}$ and ending in the out domain $%
D_{0}^{+}=D_{\phi _{2}}$. In each scattering the values of the constants of
the motion change. As this scattering is repeated again and again really $%
D_{0}$ must be considered as an infinite sequence of $D_{\phi _{i}}$}.

Then in this case, taking into account the caveat of section II.1, we can
consider that since the system is a K- system, it has a continuous spectrum,
so using the results reviewed in sections III we can say that the state $%
\rho (t)$ inside the billiard reaches an equilibrium limit $\rho _{*}$ given
by eq. (\ref{4-1}). Now, $(\omega ,p,p^{\prime }|_{\phi _{i}}$ is a
functional (i.e. a distribution or kernel) in the continuous variable $%
\omega $ but it is a trivial matrix in the discrete variables $p,$ $%
p^{\prime }$. Nevertheless, based on the observation of section II.1, we can
consider $\omega $ as a discrete variable that has been approximated by a
continuous one so we can substitute $(\omega ,p,p^{\prime }|_{\phi _{i}}$ by 
$|\omega ,p\rangle _{\phi _{i}}\langle \omega ,p^{\prime }|_{\phi _{i}}$
then the eq. (\ref{4-1}) reads 
\begin{equation}
W\lim_{t\rightarrow \infty }\rho (t)=\rho _{*}=\sum_{ip\omega }\rho (\omega
)_{\phi _{i}p}|\omega ,p\rangle _{\phi _{i}}\langle \omega ,p|_{\phi _{i}}
\label{CP.3}
\end{equation}
This is the equilibrium state of the upper part that substitutes the $\rho
(t)=|\varphi ({\bf x,}t)\rangle \langle \varphi ({\bf x,}t)|$ of eq. (\ref
{CP.1}). Now we must obtain the corresponding solution in the lower part,
solving the von Neumann equation. But this equation is linear, and now the
initial conditions are provided not by (\ref{CP.1}) but by (\ref{CP.3}), and
since in (\ref{CP.3}) $\rho _{*}$ is a linear combination of $|\omega
,p\rangle _{\phi _{i}}\langle \omega ,p^{\prime }|_{\phi _{i}},$ to obtain
the new $p_{int}$ we must only repeat the same linear combination, e.g. 
\[
\quad p_{int}=2\sum_{ip\omega }\rho (\omega )_{\phi _{i}p}%
\mathop{\rm Re}%
(\varphi _{1\omega p}({\bf x})\varphi _{2\omega p}^{*}({\bf x}))_{\phi _{i}}
\]
where 
\[
\varphi _{1\omega p}({\bf x})=\langle {\bf x|}\omega ,p\rangle _{1\phi _{i}},%
\text{ \quad }\varphi _{2\omega p}({\bf x})=\langle {\bf x|}\omega ,p\rangle
_{2\phi _{i}}
\]
are the solutions in the lower box, centered in the slits 1 and 2
respectively. Now we make the inverse transformation the one that allows to
go from eq. (\ref{4-1}) to eq. (\ref{4-2}), i.e. $|\omega ,p\rangle _{\phi
_{i}}=U_{p\phi _{i}}^{m}|\omega ,m\rangle _{\phi _{i}}$ where $U_{p\phi
_{i}}^{m}$ is the unitary transformation that diagonalizes $(\omega
,m,m^{\prime }|_{\phi _{i}}$ so 
\[
\quad p_{int}=2\sum_{ip\omega mm^{\prime }}\rho (\omega )_{\phi _{i}p}%
\mathop{\rm Re}%
[U_{p\phi _{i}}^{m}\varphi _{1\omega m}({\bf x,})(U_{p\phi _{i}}^{m^{\prime
}}\varphi _{2\omega m^{\prime }}({\bf x})^{*}]_{\phi _{i}}=
\]
\[
\quad \sum_{ip\omega mm^{\prime }}\rho (\omega )_{\omega p\phi
_{i}}[(U_{p\phi _{i}}^{m}\varphi _{1\omega m}({\bf x})(U_{p\phi
_{i}}^{m^{\prime }}\varphi _{2\omega m^{\prime }}({\bf x}))^{*}+(U_{p\phi
_{i}}^{m}\varphi _{1\omega m}({\bf x}))^{*}U_{p\phi _{i}}^{m^{\prime
}}\varphi _{2\omega m^{\prime }}({\bf x})]_{\phi _{i}}
\]
Now $\varphi _{1\omega m}({\bf x,}t)$ and $\varphi _{2\omega m^{\prime
}}^{*}({\bf x,}t)$ are eigenvalues of $H$ and $P_{x},$ and therefore also of 
$P_{y},$ then\footnote{%
Really these solutions must be added in order to satisfy the boundary
condition of the lower part of the system, but this is just another
summation that does not modify the final result.} 
\[
\varphi _{1\omega m}({\bf x}^{\prime }{\bf ,}t)\sim e^{-i\frac{{\bf m.x}%
^{\prime }}{\hbar }},\quad \varphi _{2\omega m^{\prime }}^{*}({\bf x}%
^{\prime \prime }{\bf ,}t)\sim e^{-i\frac{{\bf m}^{\prime }{\bf .x}^{\prime
\prime }}{\hbar }}
\]
But in the slits one of this functions is obtained from the other by a
displacement ${\bf s=(}s,0)$ where $s$ is the distance between the slits.
Then calling ${\bf x}^{\prime }={\bf x}-\frac{1}{2}{\bf s}$ and ${\bf x}%
^{\prime \prime }={\bf x}+\frac{1}{2}{\bf s}$ we have 
\[
\quad p_{int}=2\sum_{ipp^{\prime }\omega mm^{\prime }}\rho (\omega )_{\phi
_{i}pp^{\prime }}%
\mathop{\rm Re}%
[U_{p\phi _{i}}^{m}e^{-i\frac{{\bf m.(x-}\frac{1}{2}{\bf s)}}{\hbar }%
}(U_{p\phi _{i}}^{m^{\prime }})^{*}e^{i\frac{{\bf m}^{\prime }{\bf .(x+}%
\frac{1}{2}{\bf s)}}{\hbar }}]_{\phi _{i}}=
\]
\[
\quad 2\sum_{ipp^{\prime }\omega mm^{\prime }}\rho (\omega )_{\phi
_{i}pp^{\prime }}%
\mathop{\rm Re}%
[U_{p\phi _{i}}^{m}(U_{p\phi _{i}}^{m^{\prime }})^{*}e^{-i\frac{({\bf m-m}%
^{\prime }){\bf .x}}{\hbar }}e^{i\frac{({\bf m+m}^{\prime }).{\bf s}}{2\hbar 
}}]_{\phi _{i}}=
\]
\[
\quad \sum_{ipp^{\prime }\omega mm^{\prime }}\rho (\omega )_{\phi
_{i}pp^{\prime }}[U_{p\phi _{i}}^{m}(U_{p\phi _{i}}^{m^{\prime }})^{*}e^{-i%
\frac{({\bf m-m}^{\prime }){\bf .x}}{\hbar }}e^{i\frac{({\bf m+m}^{\prime }).%
{\bf s}}{2\hbar }}+U_{p\phi _{i}}^{m^{\prime }}(U_{p\phi _{i}}^{m})^{*}e^{i%
\frac{({\bf m-m}^{\prime }){\bf .x}}{\hbar }}e^{-i\frac{({\bf m+m}^{\prime
}).{\bf s}}{2\hbar }})_{\phi _{i}}
\]
Now we can rephrase what we have said in section II.1 but now related to the
discrete variable ${\bf m}$ instead of $\alpha _{\nu }.$ Now the number of
terms in the summation is extremely big. In fact, while in the integrable
case there will be just a few terms (see end of section IV), now the system
is not integrable and the terms may be infinite, since they arrive from
every direction\footnote{%
Each one produced by one of the scatterings that we have numerated by the $%
D_{\phi _{i}}$ of the footnote 1.}, so these ${\bf m}$ are very close. Thus
the last $\sum_{ipp^{\prime }mm^{\prime }}$ can be considered as two
integrals in the ${\bf m}$ and in the ${\bf m}^{\prime }$ that can be
changed in two integrals in the ${\bf m+}$ ${\bf m}^{\prime }$ and the ${\bf %
m-}$ ${\bf m}^{\prime }.$ In particular the integrals contain the factors $%
e^{i\frac{({\bf m-m}^{\prime }).{\bf x}}{2\hbar }},e^{-i\frac{({\bf m-m}%
^{\prime }).{\bf x}}{2\hbar }}.$ So as there is a macroscopic distance from
the two slits screen to the photographic plate ${\bf x}$ is macroscopic with
respect to $\hbar $ in such a way that we can consider that $\frac{{\bf x}}{%
\hbar }\rightarrow \infty $ and we can use the Riemann-Lebesgue theorem
concluding that 
\[
p_{int}=\sum_{ipp^{\prime \omega }mm^{\prime }}\rho (\omega )_{\phi
_{i}pp^{\prime }}[U_{p\phi _{i}}^{m}(U_{p\phi _{i}}^{m^{\prime }})^{*}e^{-i%
\frac{({\bf m-m}^{\prime }){\bf .x}}{\hbar }}e^{i\frac{({\bf m+m}^{\prime }).%
{\bf s}}{2\hbar }}+U_{p\phi _{i}}^{m^{\prime }}(U_{p\phi _{i}}^{m})^{*}e^{i%
\frac{({\bf m-m}^{\prime }){\bf .x}}{\hbar }}e^{-i\frac{({\bf m+m}^{\prime
}).{\bf s}}{2\hbar }})_{\phi _{i}}=0
\]
So the interference fringes vanish and there is decoherence in the final
equilibrium state. q. e. d.

\section{Conclusion}

1.- We have shown that the Casati-Prosen model reaches an equilibrium state
in a finite decoherence time. In this final state the interference fringes
vanish and we have decoherence. From eq. (\ref{4-3}), taking for $M$ the
electron mass and $a=1cm$ we have $t_{D}\approx 1s.$

2.- There is no environment, decoherence is produced by complexity. So the
computational result of Casati and Prosen cannot be explained by EID. But,
we have demonstrated in \cite{SIDEID} that a new combined formalism can
encompass, in a consistent way, EID and SID. In this case SID solves a
problem that cannot be solved by EID. The conclusions are that EID is a
correct theory but it is incomplete and that it can be completed with SID.

3.- All the reasoning has being done at the quantum level (with some side
remarks at the classical level) so we may say that the decoherence is
produced by quantum chaos. We will try to precise this notion based in SID
formalism in the near future.

\end{document}